\documentclass[11pt,aaspp4]{aastex}

\newcommand{\NH}{{$N_{\rm H}$}}
\newcommand{\LX}{{$L_{\rm X}$}}

\newcommand{\LHa}{$L_{{\rm H} \alpha}$}

\newcommand{\eps}{ergs s$^{-1}$}
\newcommand{\pcm}{cm$^{-2}$}

\newcommand{\asca}{{\it ASCA}}
\newcommand{\Einstein}{{\it Einstein}}
\newcommand{\rosat}{{\it ROSAT}}

\newcommand{\gtsima}{$\; \buildrel > \over \sim \;$}
\newcommand{\simgt}{\lower.5ex\hbox{\gtsima}}
\newcommand{\ltsima}{$\; \buildrel < \over \sim \;$}
\newcommand{\simlt}{\lower.5ex\hbox{\ltsima}}

\slugcomment{To appear in {\it The Astrophysical Journal.}}

\lefthead{Terashima et al.}  
\righthead{Ionizing source in LINERs}

\begin{document}

\title{Hard X-ray Emission and the Ionizing Source in LINERs}

\author{Yuichi Terashima\altaffilmark{1},
Luis C. Ho\altaffilmark{2}, and
Andrew F. Ptak\altaffilmark{3}
}

\altaffiltext{1}{NASA Goddard Space Flight Center, Code 662,
Greenbelt, MD 20771}

\altaffiltext{2}{The Observatories of the Carnegie Institution of
Washington, 813 Santa Barbara St., Pasadena, CA 91101-1292}

\altaffiltext{3}{Department of Physics, Carnegie Mellon University, 5000 Forbes
Ave., Pittsburgh, PA 15213}

\begin{abstract}

We report X-ray fluxes in the 2--10 keV band from LINERs (low-ionization 
nuclear emission-line regions) and low-luminosity Seyfert
galaxies obtained with the {\asca} satellite. Observed X-ray
luminosities are in the range between 4$\times10^{39}$ and
5$\times10^{41}$ {\eps}, which are significantly smaller than that of the
``classical'' low-luminosity Seyfert 1 galaxy NGC 4051. We found that X-ray
luminosities in 2--10 keV of LINERs with broad H$\alpha$ emission in their
optical spectra (LINER 1s) are proportional to their H$\alpha$ luminosities. 
This correlation strongly supports the hypothesis that the dominant ionizing 
source in LINER 1s is photoionization by hard photons from low-luminosity
AGNs. On the other hand, the X-ray luminosities of most LINERs without
broad H$\alpha$ emission (LINER 2s) in our sample are lower than LINER 1s at a
given H$\alpha$ luminosity. The observed X-ray luminosities in these
objects are insufficient to power their H$\alpha$ luminosities, suggesting 
that their primary ionizing source is other than an AGN, or that 
an AGN, if present, is obscured even at energies above 2 keV.

\end{abstract}


\section{Introduction}

LINERs (low-ionization nuclear emission-line regions; Heckman 1980)
are fairly common in nearby galaxies. A recent optical spectroscopic
survey of nearby galactic nuclei has shown that $\sim$30\% of bright
galaxies have LINERs (Ho, Filippenko, \& Sargent 1997a, 1997b; Ho et
al. 1997c). The ionization mechanism of LINERs is still controversial
(see Filippenko 1996 for a review). There are several ionization
mechanisms, which can explain LINER type optical emission lines, such
as (1) photoionization by low-luminosity AGNs (LLAGNs), (2)
photoionization by very hot Wolf-Rayet stars or O-stars, (3) shocks,
(4) cooling flows, and so on.

Recent multiwavelength observations argue that a significant fraction
of LINERs are genuine AGNs (e.g., Ho 1999a). Detection of an X-ray
nucleus is one of the most convincing pieces of evidence for the
presence of an AGN. From X-ray observations, the presence of LLAGNs in
LINERs is reported for several objects (Ho 1999a; Terashima et
al. 2000b, and references therein) and their luminosities ranges from
$4\times10^{39}$ to $5\times10^{41}$ {\eps}, which is about 1--3
orders of magnitude lower than typical Seyfert galaxies. If such
LLAGNs are the dominant ionizing source of the LINER-type emission
lines, the optical emission-line luminosities are expected to be
proportional to the luminosities of the ionizing photons from the
LLAGNs. Indeed, a strong positive correlation between optical
emission-line (H$\alpha$, H$\beta$, [\ion{O}{3}]) luminosities and
X-ray luminosities is observed for luminous AGNs (Kriss et al. 1982;
Elvis, Soltan, \& Keel 1984; Ward et al. 1988; Mulchaey et
al. 1994). Koratkar et al. (1995) reported that this correlation
extends to intrinsically faint objects based on {\rosat} observations
of five low-luminosity Seyfert galaxies in the soft X-ray band. A
small number of LINERs have been studied with {\Einstein}, and a
similar correlation is suggested (Halpern \& Steiner 1983; Elvis et
al.  1984).

The ionization mechanisms of LINERs, however, could be heterogeneous,
and it is also possible that ionization sources other than AGNs are
present. If other ionization mechanisms play a role, the X-ray
luminosities are expected to be small compared to given optical
emission-line luminosities (P\'erez-Olea \& Colina 1996).

In this paper, we present \asca\ measurements of X-ray fluxes of
LINERs and low-luminosity Seyfert galaxies in the 2--10 keV band and
examine the probable ionization mechanism in LINERs by comparing X-ray
luminosities with H$\alpha$ luminosities.

\section{The Data}

  We have systematically analyzed \asca\ archival and proprietary data
of LINERs in the bright galaxy sample of Ho et al. (1997a). We also
analyzed low-luminosity Seyfert galaxies (hereafter referred as
LLSeyfert) with $\log$ {\LHa} $<41$ {\eps}, which is smaller than or
comparable to that of the well-studied low-luminosity Seyfert galaxy
NGC 4051, to compare their X-ray properties with those of LINERs and
to discriminate the ionizing source in LINERs. Detailed results will
be presented in a future paper (Terashima et al. 2000b). We also use
several objects in the southern hemisphere with good \asca\ data
(Iyomoto et al. 1996, 1997). We use only galaxies for which optical
emission-line luminosities are published in the literature. Note that
our sample is not complete and that some of the galaxies in our sample
are selected based on large H$\alpha$ fluxes (see Terashima et
al. 2000a). Therefore, our sample is not an X-ray selected sample, and
so it is not significantly biased toward X-ray bright objects.

The optical classifications, adopted distances, broad and narrow
H$\alpha$ luminosities, and X-ray luminosities in the 2--10 keV band
are summarized in Table 1. We refer to objects with and without broad
H$\alpha$ emission as Type 1 (LINER 1s and LLSeyfert 1s) and Type 2
objects (LINER 2s and LLSeyfert 2s), respectively, by analogy with
Seyfert 1s and Seyfert 2s. Two objects (NGC 4192 and NGC 4569) are
classified as a transition object between a LINER and an \ion{H}{2}
nucleus. Although NGC 1097 was a LINER 2 historically
(Phillips et al. 1984), we adopt a
classification of Seyfert 1.5 based on 
Storchi-Bergmann, Baldwin, \& Wilson (1993)
since the \asca\ observation was performed after the appearance of
double-peaked broad H$\alpha$. H$\alpha$ luminosities are taken from
Ho et al. (1997a, c) except for several objects (see Table 1 for
references). We estimated the amount of reddening by the Balmer
decrement for narrow lines and corrected for reddening using the
reddening curve of Cardelli et al. (1989), where we assumed the
theoretical value of H$\alpha$/H$\beta$=3.1.  In the case that the
observed H$\alpha$/H$\beta$ is less than 3.1, no correction was
made. The reddening corrections for broad H$\alpha$ lines were made by
same amount as narrow lines.

X-ray emission is detected from all the objects except for NGC 404.
The X-ray spectra of most of the objects are represented by a
two-component model consisting of an optically-thin thermal plasma
with a temperature of $\sim$0.7 keV and a hard component. Spectra of a
few objects do not require a soft thermal component. The hard
component is well represented by a power law with a photon index of
$\sim$ 1.5--2.0. Detailed results of spectral fits are given in
Terashima et al. (2000b). The absorption column density for the hard
component ranges from $10^{20}$ to $10^{24}$ {\pcm}. We will make use
of the intrinsic luminosities of the hard component corrected for the
absorption in the 2--10 keV band. Among the objects in our sample,
seven objects (NGC 1052, 1365, 2639, 4258, 2273, 2655, and 4941) have
a heavily absorbed ({\NH}$ = 10^{23-24}$ {\pcm}) transmitted
continuum in their spectra, and the correction of the absorption is
significant. Adopted column densities of these objects are 2.0, 2.8,
4.8, 1.3, 4.6, 4.0, 10 $\times10^{23}$ {\pcm}, respectively, as
measured in Terashima et al. (2000b) and references in Table 1.
If the absorption column is greater than $10^{24}$ {\pcm} and a
transmitted hard X-ray continuum is not seen, corrections for the
absorption cannot be made. In such a case, we utilize the observed
luminosities corrected for only the Galactic absorption. These
luminosities are summarized in Table 1. The X-ray luminosities range
from 4$\times10^{39}$~{\eps} to 5$\times10^{41}$~{\eps}. Results of
{\asca} observations for several objects have been already published,
and references for them are also shown in Table 1.

\placetable{tbl-1}

\section{Results and Discussion}

\subsection{LINER 1s}

In order to examine whether optical emission lines in LINER 1s are
photonionized by high-energy photons from an AGN, we search for a
correlation between X-ray luminosities in the 2--10 keV band ({\LX})
and H$\alpha$ luminosities ({\LHa}) for objects with broad H$\alpha$
in their optical spectra. It is known that a significant positive
correlation between these two quantities exists for luminous AGNs
(e.g., Ward et al. 1988). If photoionization by an LLAGN is the
dominant ionization mechanism in LINER 1s, an {\LX}-{\LHa} correlation
is expected.  Figure 1a shows the correlation between {\LX} and {\LHa}
for LINER 1s and LLSeyfert 1s in our sample and luminous Seyfert 1s
and QSOs taken from Ward et al. (1988); the H$\alpha$ luminosities
shown represent the sum of the narrow and broad components of the
line. It is clear that the correlation extends to lower
luminosities. The same correlation using fluxes is shown in Figure 1b. 
The correlation is still significant in this plot.  This correlation
strongly suggests that the dominant ionization source in LINER 1s is
photoionization by LLAGNs with luminosities less than
$\sim5\times10^{41}$ {\eps}. The scatter in the plots could be due to
X-ray and H$\alpha$ variability of a factor of a few or more since
X-ray and H$\alpha$ luminosities were not obtained simultaneously.  A
drastic example is an appearance of a double-peaked broad H$\alpha$ in
NGC 1097, NGC 4203, and NGC 4450 (Storchi-Bergmann et al. 1993, 1995
; Shields et al. 2000; Ho et al. 2000). 
Note that the Seyfert 1.8 galaxy NGC 1365 is probably Compton thick,
and the X-ray luminosity is not corrected for absorption. Therefore,
the data point is less luminous in X-rays by about two orders of 
magnitude than the correlation.

The X-ray images in the 2--10 keV band of most LINER 1s are unresolved
within the spatial resolution of \asca\ (FWHM $\approx$ 0.5$^\prime$)
and consistent with the LLAGN interpretation as the origin of hard
X-rays. Note that hard X-ray emission from NGC 4636 is clearly
extended compared to the point-spread function, and an AGN is not the
dominant X-ray source (Matsushita et al. 1994; Terashima et
al. 2000b). Therefore, the X-ray luminosity of this object should be
regarded as an upper limit. This upper limit is still consistent with
the correlation above. If we exclude NGC 1365, 4636, and 4438, the
latter having a very small {\LX}/{\LHa} ratio ($\S$3.2), the
correlation for the sample of LINER 1s, LLSeyfert 1s, and luminous
AGNs plotted in Figure 1a is expressed as $\log${\LX} =
($0.97\pm0.07$) $\log${\LHa} + ($2.3\pm2.7$) (errors are 90\%
confidence).

\placefigure{fig-1}

\subsection{LINER 2s}

The X-ray emission from LINER 1s is most likely dominated by AGNs. If
LINER 2s are also LLAGNs, they might be obscured, by analogy with
Seyfert 2s, and the X-ray continuum emission is expected to be
absorbed by a large column density. On the other hand, with other
ionization mechanism such as hot stars or shocks resulting from starburst
activity, an {\LX}/{\LHa} value is expected to be small compared to
those for AGN (P\'erez-Olea \& Colina 1996). In Figure 2, we compare
the luminosity ratio {\LX}/{\LHa} for the different object classes to
examine whether the ionization mechanism in LINER 2s is the same as in
LINER 1s and LLSeyferts; in this comparison we make use only of the
narrow component of the H$\alpha$ line. The mean of the {\LX}/{\LHa}
distributions for LLSeyfert 1s, LLSeyfert 2s, LINER 1s, and LINER 2s
are 52.8, 10.3, 67.3, and 7.7, respectively. Thus the mean of
{\LX}/{\LHa} for LINER 2s is about one order of magnitude smaller than
those for LLSeyfert 1s and LINER 1s. A few LINER 2s (NGC 4261, NGC
4594, and NGC 4736) have an {\LX}/{\LHa} ratio similar to LINER 1s and
LLSeyfert 1s, while many LINER 2s have a small {\LX}/{\LHa}. Some
LLSeyferts with small {\LX}/{\LHa} are most probably Compton thick
objects in which only scattered X-rays are observed (NGC 1365, Iyomoto
et al. 1997; NGC 1386, Maiolino et al. 1998, Bassani et al. 1999; NGC
5194, Terashima et al. 1998b) since these objects have a strong Fe-K
emission line with an equivalent width of $>$ 1 keV, and the intrinsic
{\LX}/{\LHa} values could be 1--2 orders of magnitude larger than
observed. There is no clear X-ray evidence for the presence of heavily
obscured AGN in NGC 3079 and NGC 7743, although such a possibility
cannot be ruled out.

\placefigure{fig-2}

If we assume a spectral energy distribution from UV to X-ray is
power-law shape with an index of --1 ($f_{\nu} \propto \nu^{-1}$) (Ho
1999b), $\log$ {\LX}/{\LHa} should be greater than 1.4 to drive
H$\alpha$ luminosities by photoionization under Case B recombination
and a covering fraction of unity (Osterbrock 1989). Since X-ray
luminosities of most LINER 2s are small ({\LX}$< 2\times10^{40}$
\eps), X-ray binaries in the host galaxy might also contribute to
observed X-ray fluxes. Actually some objects such as NGC 3607, NGC
4111, NGC 4374, and NGC 4569 have hard band ($>$2 keV) images extended
to several kpc (Terashima et al. 2000a, b), which implies a lower
value of {\LX}/{\LHa} for the nuclear component. Therefore the objects
with low $\log$ {\LX}/{\LHa} ($<$1) are too X-ray weak to ionize
optical emission lines, even if X-ray variability of a factor of a few
is also taken into account. If an AGN is present and is the dominant
ionizing source in these objects, it should be obscured even at
energies above 2 keV. Although no clear evidence for the presence of
heavily obscured AGN (e.g. heavily absorbed X-ray continuum and/or
strong Fe-K emission line) has been obtained so far, the current data
cannot rule out such a possibility. Alternatively, there might be
other ionization sources. The {\it Hubble Space Telescope} UV spectra
of NGC 404 and NGC 4569 actually show large number of hot stars
concentrated in pc-scale nuclear regions (Maoz et al. 1998), and these
objects could be examples of LINERs ionized by stellar sources
(Terashima et al. 2000a).

\subsection{Fraction of AGNs in Bright Galaxies}

According to the optical spectroscopic survey by Ho et al. (1997a, b),
Seyferts, LINER 1s and LINER 2s are detected in 11\%, 5\%, 28\% of
northern ($\delta>0^{\circ}$) bright ($B_T\leq12.5$ mag) galaxies. If
we assume an extreme case that all Seyferts and LINER 1s are AGNs and
all LINER 2s are not AGNs, the fraction of AGN is estimated to be 16\%
of bright galaxies.  This percentage, however, should be regarded as a
lower limit because extremely weak broad H$\alpha$ is difficult to
detect unambiguously (Ho et al. 1997c).  Furthermore, some LINER 2s
clearly indicate AGN-like activity. For example, NGC 4261 has
prominent radio jets and kinematic evidence for a massive black hole
(Ferrarese, Ford, \& Jaffe 1996). Other examples include the
``Sombrero'' galaxy (NGC 4594) and NGC 4736. The {\LX}/{\LHa} values
for these object are similar to LINER 1s and LLSeyferts. This fact
also indicates that the {\LX}/{\LHa} ratio is a good indicator of the
presence of AGNs. Hard X-ray surveys conducted at high angular
resolution, such as afforded by {\it Chandra}, would be crucial to
refine the true AGN fraction in nearby galaxies.

\acknowledgments

The authors are grateful to all the {\asca} team members. YT is
supported by the Japan Society for the Promotion of Science
Postdoctoral Fellowships for Research Abroad. The research of LCH is
partially supported by NASA grant NAG 5-3556 and by NASA grants
GO-06837.01-95A and AR-07527.02-96A from the Space Telescope Science
Institute (operated by AURA, Inc., under NASA contract
NAS5-26555). AFP is partially supported by NASA grant NAG 5-8093.

\newpage

\begin{deluxetable}{lcccccccc}
%\tabletypesize{\scriptsize}
\tablecaption{X-ray and H$\alpha$ luminosities for observed galaxies
\label{tbl-1}
}
\tablewidth{0pt}
\tablehead{
\colhead{Name}	& \colhead{Distance}	&\colhead{ Class}	& \colhead{$\log L$(H$\alpha$)} & \colhead{$\log L$(H$\alpha$)} 	& \colhead{$\log L_{\rm X}$} 	& \colhead{$\log${\LX}/{\LHa}}		& \multicolumn{2}{c}{reference}\\ 
	& 	         &			& (Broad)	& (Narrow)	& (2--10 keV)	&  	& X	& H$\alpha$\\
	& [Mpc]		&			& [{\eps}]	& [{\eps}]	& [{\eps}]
}
\startdata
%LINER 1.9\\
NGC 315 & 65.8 & L1.9& 39.92 & 39.61 & 41.70  & 2.09	& 1	& \\
NGC 1052 & 17.8 & L1.9& 39.54 & 39.45 & 41.58 & 2.13	& 2,3	& 1,2\\
NGC 3998 & 21.6 & L1.9& 40.59 & 40.43 & 41.67 & 1.24	& 4	& \\
NGC 4203 & 9.7 & L1.9& 38.59 & 38.34 & 40.40  & 2.06	& 5\\
NGC 4438 & 16.8 & L1.9& 39.54 & 40.04 & 39.96 & --0.08	&\\
NGC 4450 & 16.8 & L1.9& 38.48 & 38.51 & 40.34 & 1.83	&	& 1,2\\
NGC 4579 & 16.8 & S1.9/L1.9 & 39.49 & 39.53 & 41.18 & 1.65 & 4,6\\
NGC 4636 & 17.0 & L1.9& 38.37 & 38.27 & 40.22 & 1.95	& 7--11\\
NGC 5005 & 21.3 & L1.9& 40.05 & 39.48 & 40.59 & 1.11	& \\
\tableline
%LINER 2\\
NGC 404 & 2.4 & L2& ... & 37.82 & $<$ 37.66   & $<-$0.16	& 12\\
NGC 3507 & 19.8 & L2& ... & 39.60 & 39.92 & 0.32 	& \\
NGC 3607 & 19.9 & L2& ... & 39.53 & 40.16 & 0.63 	& \\
NGC 4111 & 17.0 & L2& ... & 39.85 & 39.94 & 0.09 	& 12\\
NGC 4192 & 16.8 & T2& ... & 40.49 & 39.58 & --0.91	& 12\\
NGC 4261 & 35.1 & L2& ... & 39.82 & 41.18 & 1.36 	& 1, 13\\
NGC 4374 & 16.8 & L2& ... & 39.35 & 40.33 & 0.98 	& 8--10\\
NGC 4457 & 17.4 & L2& ... & 39.79 & 39.98 & 0.20 	& 12\\
NGC 4569 & 16.8 & T2& ... & 40.66 & 40.08 & --0.58	& 12	& 3\\
NGC 4594 & 20.0 & L2& ... & 39.82 & 41.00 & 1.18 	& 4,14\\
NGC 4736 & 4.3 & L2& ... & 38.12 & 39.64  & 1.52 	& 15	& 4\\
NGC 7217 & 16.0 & L2& ... & 39.78 & 39.86 & 0.08 	& \\
\tableline
%Seyfert 1--1.9\\
NGC 1097 & 14.5 & S1.5& 40.55 & 39.53 & 40.63 & 1.10 	& 16	& 5\\
NGC 1365 & 16.9 & S1.8& 42.02 & 41.08 & 40.50 & --0.58	& 17	& 6\\
NGC 2639 & 42.6 & S1.9& 39.79 & 40.48 & 41.66 & 1.18 	& 18	& 1,2\\
NGC 3031 & 1.4 & S1.5& 38.41 & 38.44 & 39.60  & 1.16 	& 19\\
NGC 4258 & 6.8 & S1.9& 38.90 & 38.60 & 40.83  & 2.23 	& 4,20\\
NGC 4565 & 9.7 & S1.9& 38.38 & 38.46 & 39.77  & 1.31 	& 21\\
NGC 4639 & 16.8 & S1.0& 39.75 & 38.39 & 40.54 & 2.15 	& 22\\
NGC 5033 & 18.7 & S1.5& 40.32 & 39.70 & 41.37 & 1.67 	& 23\\
\tableline
%Seyfert 2\\
NGC 1386 & 16.9 & S2& ... & 40.94 & 40.34 & --0.60	& 17	& 7\\
NGC 2273 & 28.4 & S2& ... & 41.00 & 41.93 & 0.93 	&\\
NGC 2655 & 24.4 &S2 & ... & 40.04 & 41.46 & 1.42 	&\\
NGC 3079 & 20.4 & S2& ... & 41.04 & 40.40 & --0.64	&4,24\\
NGC 3147 & 40.9 & S2& ... & 40.02 & 41.53 & 1.51 	& 4,25\\
NGC 4501 & 16.8 & S2& ... & 39.08 & 40.30 & 1.22 	& \\
NGC 4941 & 6.4 & S2& ... & 39.81 & 40.69  & 0.87 	& 	& 7\\
NGC 5194 & 7.7 & S2& ... & 40.00 & 39.86  & --0.14	& 4,26	& 3\\
NGC 7743 & 24.4 & S2& ... & 40.29 & 39.57 & --0.72	& \\
\enddata
%\tablenum{1}
\tablerefs{X-ray luminosities: (1) Matsumoto et al. 2000; 
(2) Guainazzi \& Antonelli  1999;(3)Weaver et al. 1999; 
(4) Ptak et al. 1999; (5) Iyomoto et al. 1998; 
(6) Terashima et al. 1998a; (7) Awaki et al. 1994; 
(8) Matsushita et al. 1994; (9) Matsumoto et al. 1996; 
(10) Boute \& Fabian 1998; (11) Allen, Di Matteo, \& Fabian 1999; 
(12) Terashima et al. 2000a; (13) Sambruna et al. 1999; (14) 
Nicholson et al. 1998; (15) Roberts et al. 1999; (16) Iyomoto et al. 
1996; (17) Iyomoto et al. 1997; (18) Wilson et al. 1998; (19) 
Ishisaki et al. 1996; (20) Makishima et al. 1994; (21) Mizuno et al. 1999; 
(22) Ho et al. 1999; (23) Terashima et al. 1999; 
(24) Dahlem et al. 1998; (25) Ptak et al. 1996; (26) Terashima et al. 1998b\\
H$\alpha$ luminosities: (1) L.~C. Ho 2000, private communication; (2) Ho et al.
1997b; (3) Keel 1983; (4) Taniguchi et al. 1996; (5) Storchi-Bergmann 
et al. 1993; (6) V\'eron-Cetty \& V\'eron 1986; (7) Storchi-Bergmann \& 
Pastriza 1989; others Ho et al. 1997a, b
}
\end{deluxetable}

\clearpage

\begin{center}
{\bf Figure Captions}
\end{center}

\figcaption[]{
({\it a}) Correlation between X-ray and H$\alpha$ luminosities for LINER 1s,
low-luminosity Seyfert 1s, and luminous type 1 AGNs taken from Ward et al. 
(1998).,
({\it b}) Correlation between X-ray and H$\alpha$ fluxes for the same sample
as ({\it a}).  The triangle and the cross correspond to NGC 1365 and NGC 4438, 
respectively.
\label{fig-1}
}

\figcaption[]{
$\log$ {\LX}/{\LHa} for low-luminosity Seyfert 1s, 
low-luminosity Seyfert 2s, LINER 1s, and LINER 2s. 
The symbol "$<$" denotes an upper limit.
\label{fig-2}
}

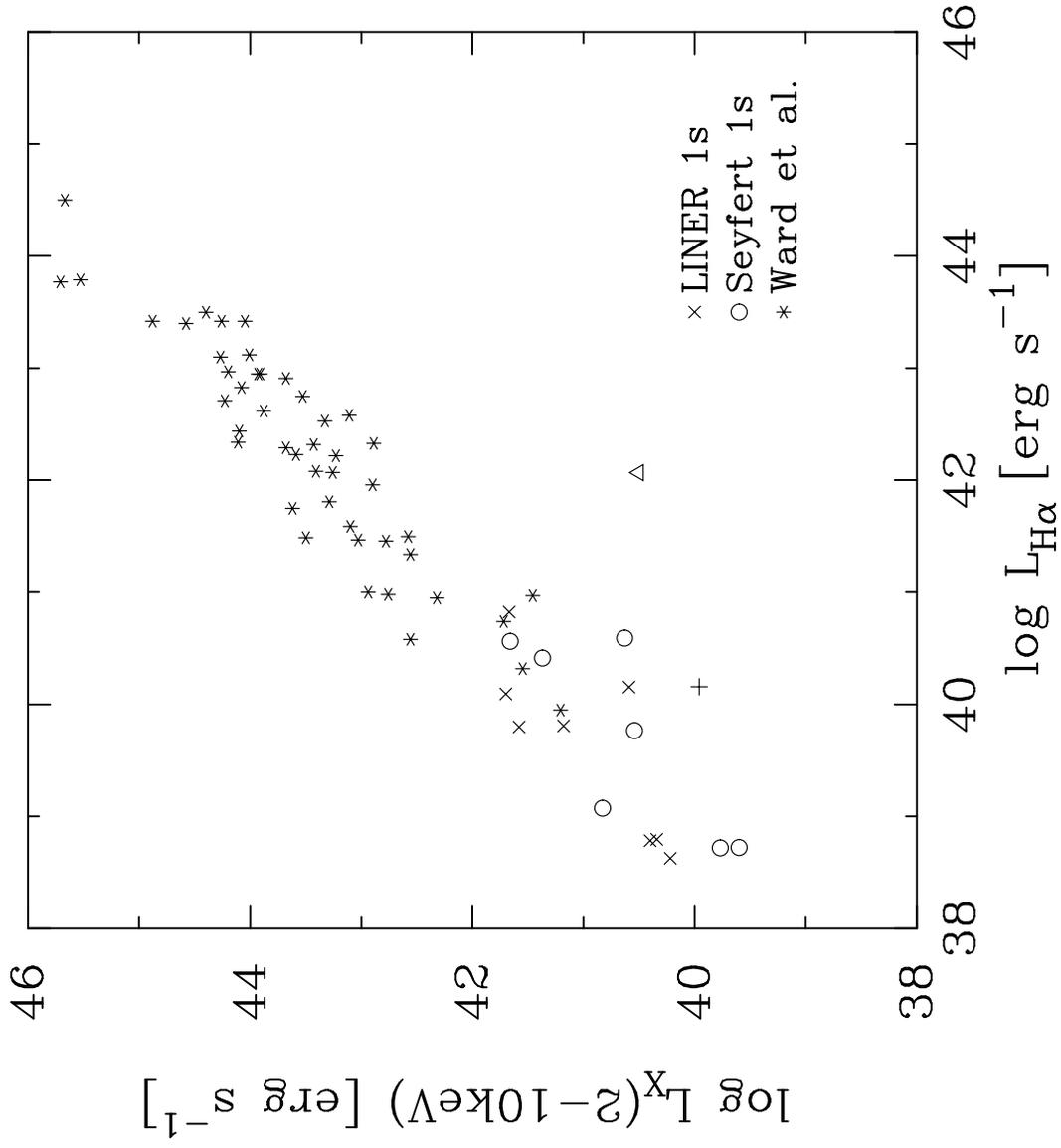
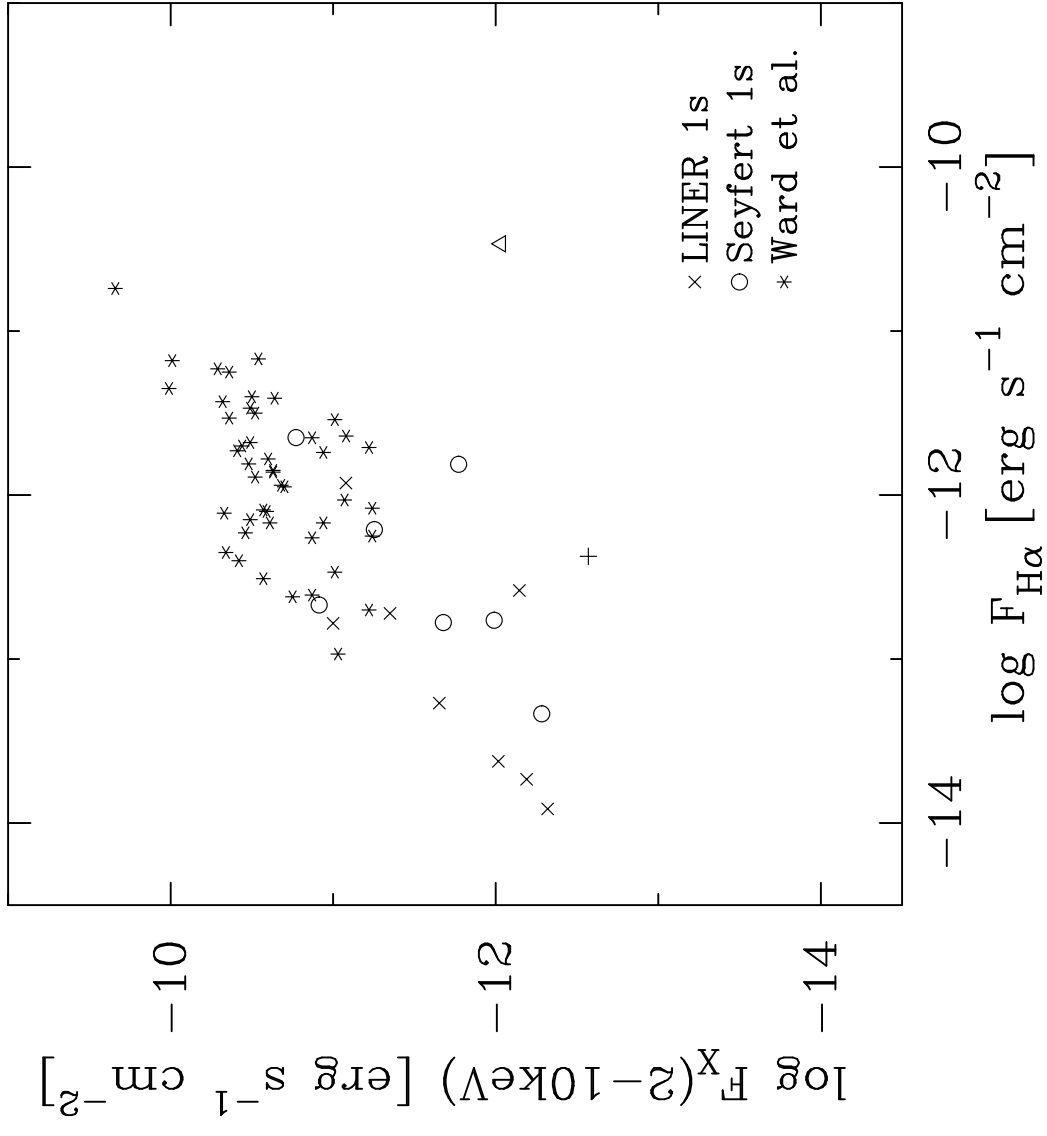
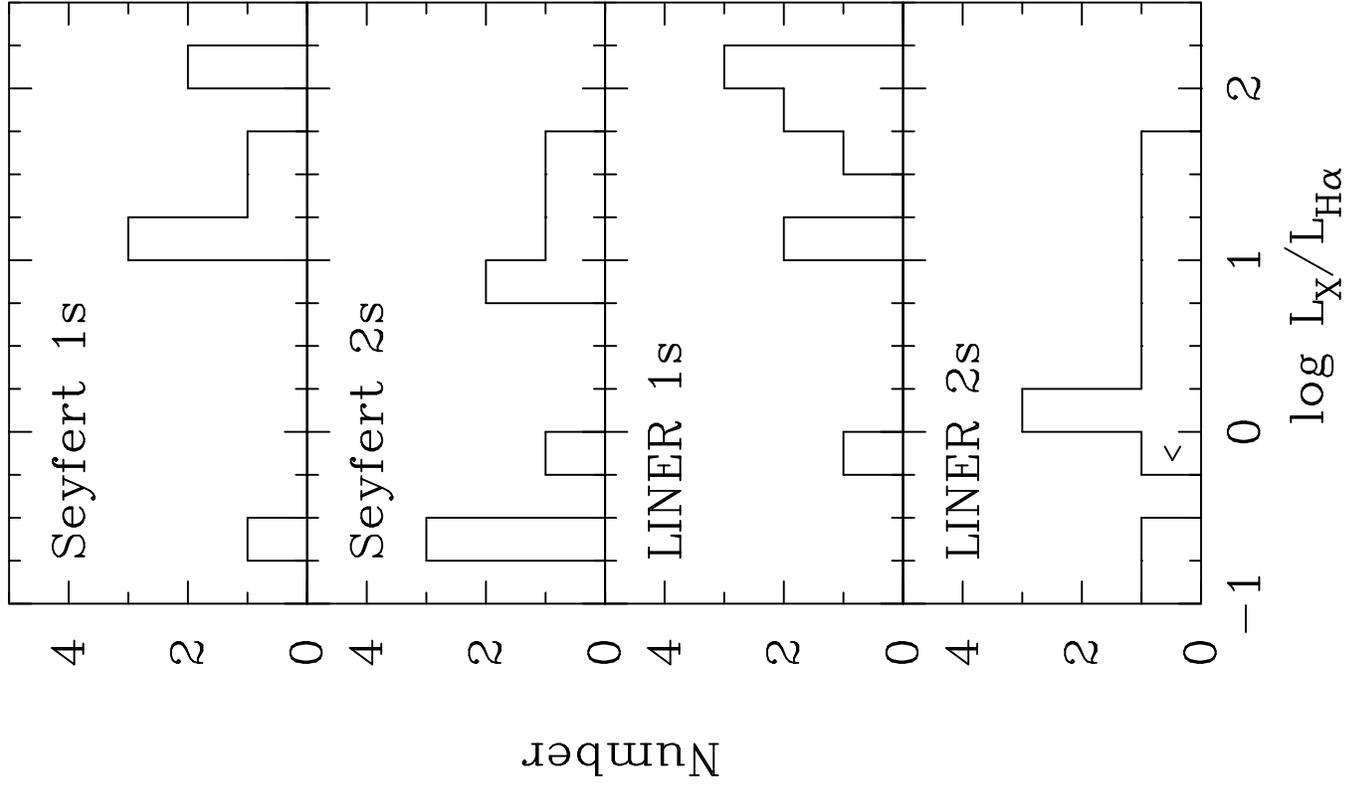
\end{document}